\documentclass[twocolumn,preprintnumbers,amsmath,amssymb,bibnotes]{revtex4}

\usepackage[usenames]{color}
\usepackage{graphicx}
\usepackage{float}

\begin{document}
\title{Sisyphus Cooling of Electrically Trapped Polyatomic Molecules}
\author{M. Zeppenfeld}
\email{martin.zeppenfeld@mpq.mpg.de}
\author{B.G.U. Englert}
\author{R. Gl\"ockner}
\author{A. Prehn}
\author{M. Mielenz}
\altaffiliation[Present address:]{ Albert-Ludwigs-Universit\"at Freiburg, Physikalisches Institut, 
Hermann-Herder-Str. 3, D-79104 Freiburg, Germany}
\author{C. Sommer}
\altaffiliation[Present address:]{ Institute for Molecular Science, National Institutes of Natural Sciences, Myodaiji, Okazaki 444-8585, Japan}
\author{L.D. van Buuren}
\altaffiliation[Present address:]{ Radiotherapy Department, The Netherlands Cancer Institute, Antoni van Leeuwenhoekhuis, 1066 CX Amsterdam, the Netherlands}
\author{M. Motsch}
\altaffiliation[Present address:]{ Laboratorium f{\"u}r Physikalische Chemie, ETH Z{\"u}rich, CH-8093 Z{\"u}rich, Switzerland}
\author{G. Rempe}
\affiliation{Max-Planck-Institut f\"ur Quantenoptik, Hans-Kopfermann-Str. 1, D-85748 Garching, Germany}
\date{\today}

%\tableofcontents

\maketitle

{\bf
The rich internal structure and long-range dipole-dipole interactions establish polar molecules as unique instruments for quantum-controlled applications and fundamental investigations. Their potential fully unfolds at ultracold temperatures, where a plethora of effects is predicted in many-body physics~\cite{Goral02,Micheli06}, quantum information science~\cite{DeMille02,Andre06}, ultracold chemistry~\cite{Krems08,Zuchowski09}, and physics beyond the standard model~\cite{Hinds97,Hudson11}. These objectives have inspired the development of a wide range of methods to produce cold molecular ensembles~\cite{Weinstein98,Gupta99,Bethlem99,Junglen04,Fulton04,Sommer09}. However, cooling polyatomic molecules to ultracold temperatures has until now seemed intractable. Here we report on the experimental realization of opto-electrical cooling~\cite{Zeppenfeld09}, a paradigm-changing cooling and accumulation method for polar molecules. Its key attribute is the removal of a large fraction of a molecule's kinetic energy in each step of the cooling cycle via a Sisyphus effect, allowing cooling with only few dissipative decay processes. We demonstrate its potential by reducing the temperature of about $10^6$ trapped CH$_3$F molecules by a factor of $13.5$, with the phase-space density increased by a factor of $29$ or a factor of $70$ discounting trap losses. In contrast to other cooling mechanisms, our scheme proceeds in a trap, cools in all three dimensions, and works for a large variety of polar molecules. With no fundamental temperature limit anticipated down to the photon-recoil temperature in the nanokelvin range, our method eliminates the primary hurdle in producing ultracold polyatomic molecules. The low temperatures, large molecule numbers and long trapping times up to $27$\,s will allow an interaction-dominated regime to be attained, enabling collision studies and investigation of evaporative cooling toward a BEC of polyatomic molecules.}

%%%%%%%%%%%%%%%%%%%%%%%%%%%%
The ability to prepare ultracold molecular ensembles has an application potential akin to that of ultracold atoms some decades ago. In fact, the association of KRb dimers~\cite{Ni08} as well as the laser cooling of SrF~\cite{Shuman10} has brought fascinating physics within reach. However, both approaches are restricted to a highly specialized set of purely diatomic molecule species. In order to investigate fundamental physics based on relativistic effects near heavy nuclei or parity violation effects in chiral molecules, or to study molecules of astrophysical, biological, or chemical interest, a more general approach to preparing ultracold molecular ensemble is imperative. This holds in particular for the rich chemical variety of carbon-, nitrogen-, or oxygen-based molecules for which the constituent atoms have not even been laser cooled. Devising a dissipative process to cool such molecules into the ultracold regime has been an exceedingly challenging problem. The standard approach for atoms, laser cooling, is in general impossible for molecules due to the lack of suitable cycling transitions. Creating an artificial cycling transition via cavity cooling~\cite{Vuletic00} has not been demonstrated despite substantial experimental~\cite{Motsch10} and theoretical~\cite{Nagy06,Zeppenfeld10,Xuereb11} effort. Likewise, evaporative or sympathetic cooling to ultracold temperatures~\cite{Lara06} has not been realized due to lack of density or losses from inelastic collisions.

A particularly promising general framework to cool molecular ensembles is to replace the weak photon recoil in laser cooling with sufficiently strong forces to remove the entire molecule's kinetic energy in a single process~\cite{Zeppenfeld09,Narevicius09}. A first step towards this approach is the recent achievement of accumulation of NH molecules using a single-photon transition~\cite{Riedel11}. Here we present a full implementation of opto-electrical cooling~\cite{Zeppenfeld09} for molecules stored in an electric trap, featuring both accumulation and cooling. Energy is extracted by allowing molecules to move up and down an electric field gradient in different states with differing Stark energies while dissipation to remove entropy is provided by a spontaneous vibrational decay. The scheme is applicable to all molecules with strong electric field interaction and pure ro-vibrational states, and thus constitutes a general method for cooling molecules to ultracold temperatures.

\begin{figure*}[ht]
\centering
\includegraphics[width=1\textwidth]{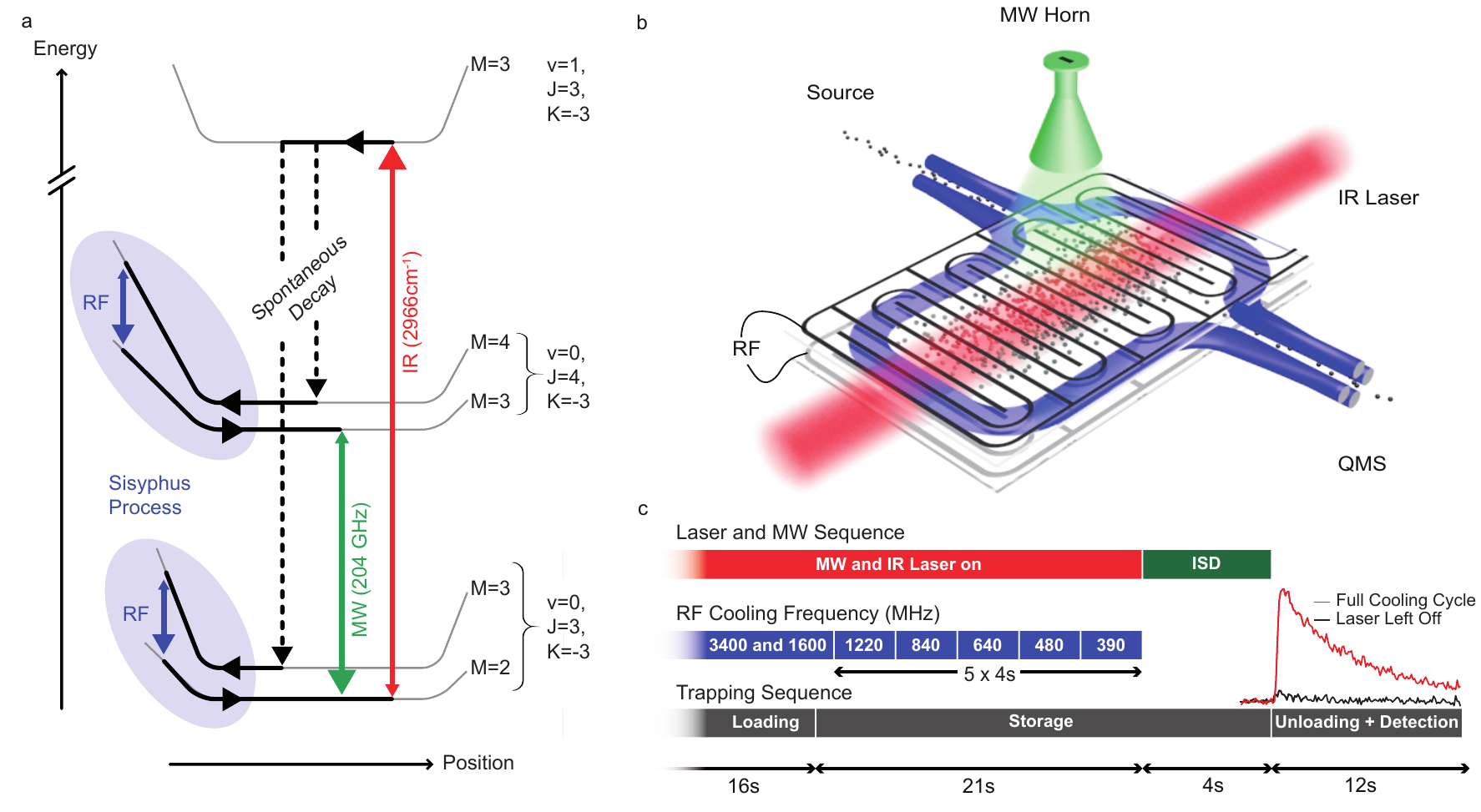}
\caption{{\bf Implementation of molecule cooling.} {\bf a,} Radiation couples molecular states which experience a position-dependent potential energy in an electric trap as indicated. Molecules following the emphasized route lose more kinetic energy when entering the strong-field edge region of the trap than they regain when returning to the trap centre in a more weakly trapped state. Combined with the unidirectionality of the optical pumping back to the strongly trapped states, this leads to cooling. {\bf b,} During cooling the molecules are confined in a microstructured electric trap~\cite{Englert11}, with the IR, MW, and RF fields applied as indicated. {\bf c,} Experimental sequence for cooling: warm molecules are continuously loaded into the trap for $16$\,s with the IR, MW, and two RF fields at $3.4$\,GHz and $1.6$\,GHz already applied. The trap is then electrically closed~\cite{Englert11} and the RF frequency is decreased stepwise to $390$\,MHz over the next $21$\,s. Before unloading the molecules for detection, an internal-state-discrimination (ISD) technique is applied during $4$ additional seconds (see methods).}\label{scheme}
\end{figure*}

Reducing the temperature by a large factor requires the cooling cycle to be repeated, and consequently, control being maintained over the internal molecular state. This is achieved using the level scheme shown in Fig.~\ref{scheme}a. An electric trap as depicted in Fig.~\ref{scheme}b with a homogeneous electric field in the trap centre and strongly increasing fields near the trap boundary~\cite{Englert11} leads to a position-dependent potential energy for the relevant molecular states as shown. We label states with vibrational quantum number $v$ and symmetric-top rotational quantum numbers $J, K, M$~\cite{moleculetheory} as $|v;J,\mp K,\pm M\rangle$ with $\mp K$ chosen positive. For a parallel vibrational transition with $\Delta K=0$, the state $|1;3,3,3\rangle$ decays to the four rotational $v=0$ states shown, with decay to the $|0;4,3,2\rangle$ state ignored due to the small Clebsch-Gordan coefficient of $\frac{1}{144}$. Coupling the more weakly trapped $|0;3,3,2\rangle$ and $|0;4,3,3\rangle$ states with microwave (MW) radiation and driving the $|0;3,3,2\rangle$ to $|1;3,3,3\rangle$ transition with an infrared (IR) laser results in optical pumping to the strongly trapped $|0;3,3,3\rangle$ and $|0;4,3,4\rangle$ states. Adding a radio-frequency (RF) to couple neighbouring $M$-sublevels in strong electric fields completes the opto-electrical cooling cycle. Losses to the untrapped $M=0$ states are avoided by coupling the neighbouring $M$-sublevels with the RF at a rate which is slow compared to the optical pumping~\cite{support}. Note that Stark detuning due to the electric fields plays a key role in selectively addressing only the desired IR and MW transitions~\cite{support}.

The experimental sequence to prepare and detect a sample of cooled molecules is shown and explained in Fig.~\ref{scheme}c. In addition to cooling, we apply the following two experimental sequences. First, a reference no-cooling measurement with the IR, MW, and RF fields left off and the storage time shortened by $20$\,s results in an uncooled molecular ensemble. Second, reintroduction of the IR and MW fields results in accumulation of molecules, also without cooling. Here, molecules entering the trap in the weakly trapped states $|0;3,3,M\rangle$ and $|0;4,3,M\rangle$ with $1\le M\le J-1$ are pumped to the strongly trapped $|0;3,3,3\rangle$ and $|0;4,3,4\rangle$ states.

\begin{figure*}[t]
\centering
\includegraphics{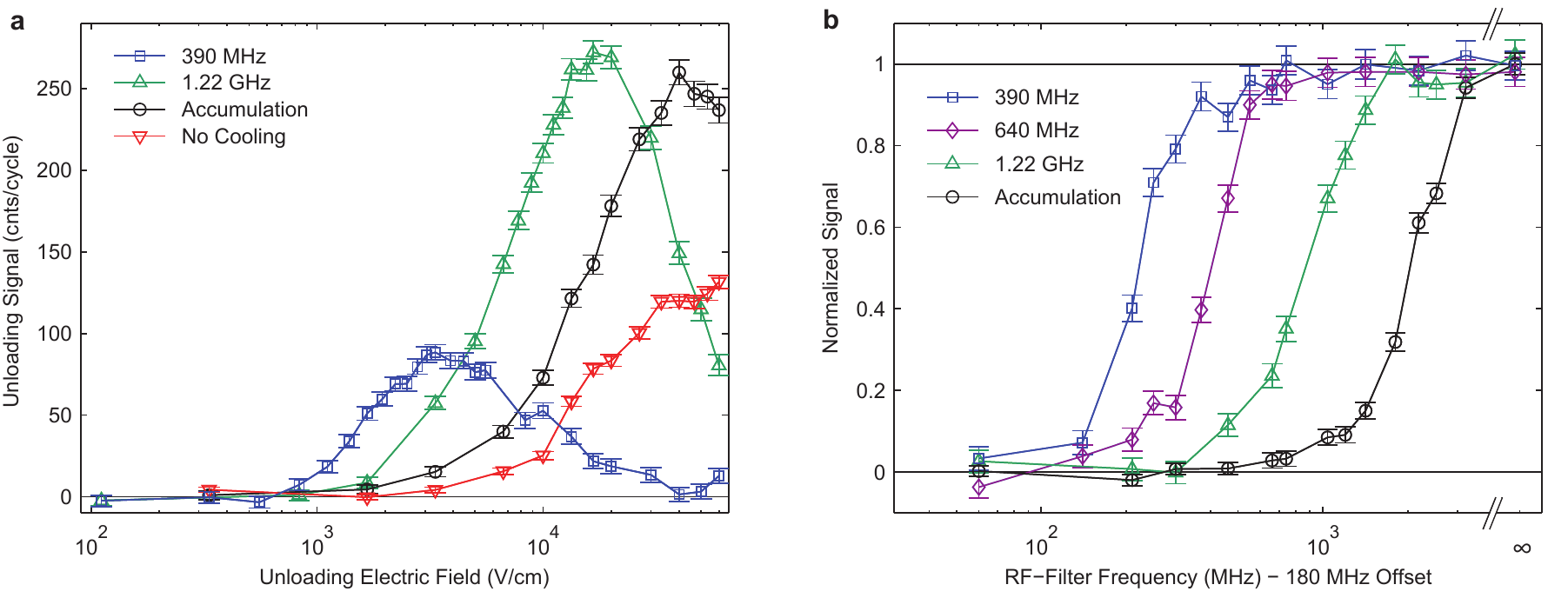}
\caption{{\bf Demonstration of opto-electrical cooling.} {\bf a,} Detected molecule signal vs.\ trap electric field strength during unloading. In addition to results for the no-cooling measurement, the accumulation, and the full cooling cycle to $390$\,MHz, we show results for cooling to the intermediate frequency of $1.22$\,GHz. Cooling results in a more than 10-fold decrease in the optimal unloading voltage and, together with the accumulation, in a multifold signal increase at low unloading voltages. {\bf b,} Relative unloading signal with RF as a knife-edge filter applied. The legend shows the RF cooling frequency (except for accumulation) after which the much stronger RF knife is applied. We subtract $180$\,MHz from the RF knife frequency to account for the offset electric field at the centre of the trap.}\label{filters}
\end{figure*}

All experimental sequences end by unloading the molecules from the trap and guiding them to the ionization volume of a quadrupole mass spectrometer (QMS) for detection. This results in signals as shown in the inset to Fig.~\ref{scheme}c, in this case for the full cooling cycle to $390$\,MHz as well as for the same cycle but with the IR laser left off. Remarkably, the sequence with laser on results in a strong molecule signal whereas leaving the laser off eliminates the signal almost entirely. Thus, the excitation by the laser and the subsequent spontaneous decay clearly plays a key role in keeping molecules in the trap, indicating that a key element of the cooling cycle is working. Note that state-discriminating measurements (see methods) show that $80$\,\% of the molecules with laser on are in the states $|0;3,3,M\rangle$ and $|0;4,3,M\rangle$ with most of these in the top $M$-sublevels~\cite{support}, establishing our experiment as a practically state-selected source of molecules. This compares to about $20$\,\% of molecules initially entering the trap in the states $|0;3,3,M\rangle$ and $|0;4,3,M\rangle$ potentially involved in the cooling.

For proof of cooling, we now present four complementary measurements. First, we examine the effect of the trap electric field strength on the molecules during unloading. The field strength during this time is particularly important since it determines both the confining potential for the molecules and the molecule transport efficiency to the detection QMS. In particular, a lower electric field strength reduces the trap depth so that hotter molecules are lost, and a higher electric field strength during unloading inhibits colder molecules from finding the trap exit hole to the QMS. We thus expect that the molecule signal falls off in both directions from a maximum at a confining electric field strength which is proportional to the temperature of the molecules, depending on the molecule Stark shifts. Fig.~\ref{filters}a shows the measured molecule signal vs.\ unloading electric field strength for four different experimental cycles. Application of the cooling sequence results in a strong decrease in the optimal unloading electric field strength, from about $45$\,kV/cm to $16$\,kV/cm and $3.3$\,kV/cm. Since the plotted signal is mainly from molecules in the states $|0;3,3,3\rangle$ and $|0;4,3,4\rangle$~\cite{support} which have a similar Stark shift, our data clearly shows that a large amount of cooling takes place. The relatively low signal for the full cooling cycle is to a large part due to trap losses during the additional $20$\,s of storage compared to accumulation and no-cooling, accounting for a $2.4$ times smaller signal (see below). Note that the accumulation alone results in approximately a 2-fold larger molecules signal, thus proving an increase in the phase-space density already here.

As a second measurement, we use a strong RF field as a knife to drive transitions to untrapped states and thus eliminate hotter molecules from the trap. Results are shown in Fig.~\ref{filters}b. For all three measurements with cooling, the cooling cycle is completed to $390$\,MHz. However, two additional seconds of storage are inserted immediately after the indicated cooling frequency for application of the RF knife. This allows for practically identical experimental sequences. Only for accumulation, the molecules are unloaded directly. We observe a sharp flank in the molecule signal at a knife frequency which strongly depends on the frequency of the cooling step just before application of the knife. The position of the flank is shifted by approximately a factor of $10$, similar to the shift observed in Fig.~\ref{filters}a. Again, the dramatic shift can only be explained by a large amount of cooling. Note that the flanks of the four measurements are well separated, indicating that the energy distributions of the molecules in the four measurements hardly overlap.

Both of the previous measurements are sensitive to the internal molecular state since Stark shift and Stark splitting directly determine the effect of the trap voltages and RF frequencies, respectively. To avoid this effect, we perform a third experiment and directly measure the velocities of the molecules via a time-of-flight technique. Specifically, we switch the $55$\,cm long quadrupole guide segment leading to the QMS on and off and measure the arrival time of the molecules. Note that the unloading voltages affect the distribution of molecules arriving at the QMS, with colder (hotter) molecules being preferentially detected at lower (higher) voltages~\cite{support}. However, by measuring at the optimal unloading voltages, the effect will be the same for all measurements relative to the ensemble in the trap. We use the same four experimental
\begin{figure}[H]
\includegraphics[width=.47\textwidth]{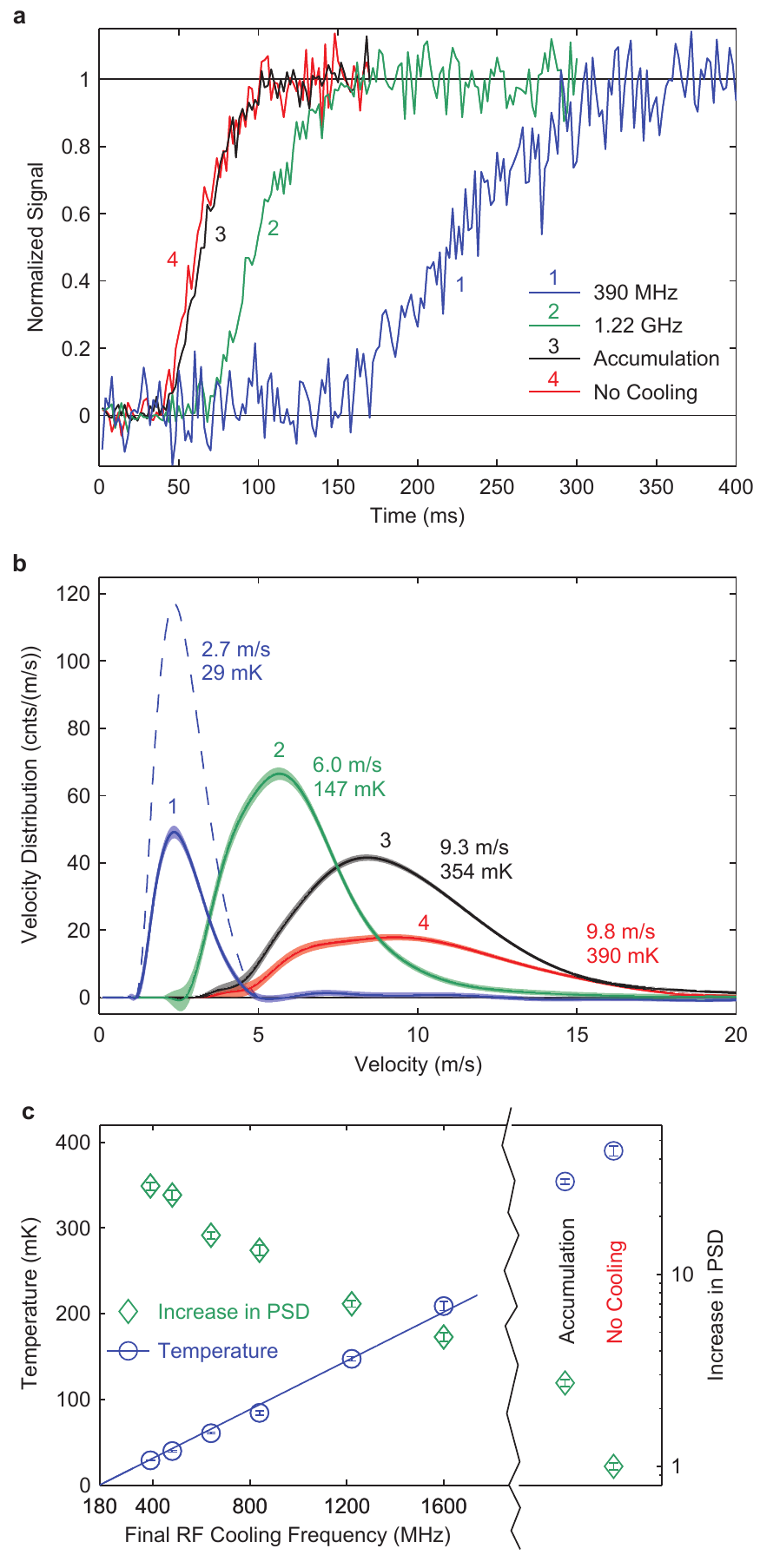}
\caption{{\bf Time-of-flight measurements to determine the molecule velocities.} {\bf a,} Rising edge of the normalized trap unloading signal for the same variations of the experimental sequence as in Fig.~\ref{filters}a. For cooled molecules one needs to wait over three times longer before molecules arrive at the QMS. {\bf b,} Molecular velocity distributions derived from the rising-edge signals shown in ({\bf a}) taking into account the number of detected molecules. The dashed blue curve shows the hypothetical result if trap losses (shown in Fig.~\ref{lifetime}) could be eliminated. The smooth appearance of the data is due to convolution with a Gaussian. The shaded regions represent the $1\sigma$ statistical error. We include the mean velocity and temperature calculated from the data sets. {\bf c,} Measured molecule temperature and total change in phase-space density (PSD) after each intermediate RF frequency as well as for no-cooling and accumulation.}\label{velocity}
\end{figure}
\noindent
sequences as for Fig.~\ref{filters}a with the results shown in Fig.~\ref{velocity}. Calculating the mean longitudinal velocity $\langle v_z\rangle$ from the data and deriving the molecular temperature $T$ via $\frac{1}{2}k_BT=\frac{1}{2}m\langle v_z\rangle^2$ we arrive at the central result: the molecules have been cooled by a factor $13.5$ from $390$\,mK to $29$\,mK. Together with the estimated molecule density, this corresponds to a factor $29$ increase in phase-space density. In fact, the cooling and accumulation by themselves increase the phase-space density by a factor of $70$, but this is reduced by trap losses during cooling.

\begin{figure}[t]
\centering
\includegraphics{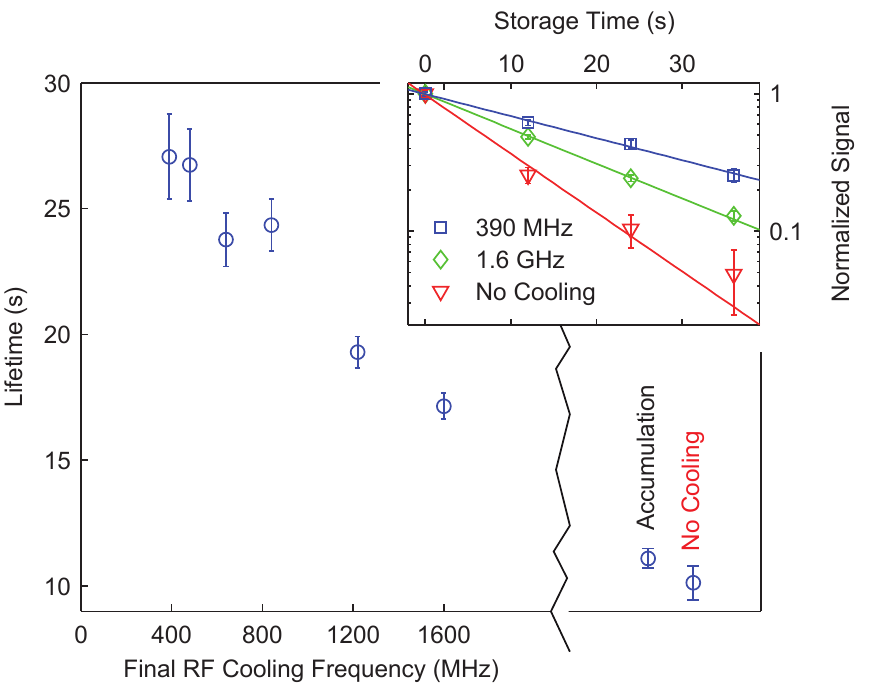}
\caption{{\bf Trap lifetime for cooled and uncooled molecules.} Cooling substantially increases the trap lifetime. The inset shows the raw data for three of the lifetime measurements.}\label{lifetime}
\end{figure}

As a fourth and final measurement, Fig.~\ref{lifetime} shows the lifetime of molecules in the trap as a function of the final cooling frequency. As expected for cooling~\cite{Englert11}, the lifetime increases from about $10$\,s without cooling to about $27$\,s with maximal cooling. This dramatic change confirms once more that cooling takes place. By integrating the measured loss rate over time, we find that the additional $20$\,s of storage for cooling to $390$\,MHz compared to no-cooling or accumulation leads to a loss of about $60$\,\% of the molecules, as has been used above in the discussion of Fig.~\ref{filters}a and in Fig.~\ref{velocity}b. For better trapping or faster cooling, this loss could be reduced.

In principle, opto-electrical cooling can be continued to much lower temperatures. However, losses during cooling and the low detection efficiency make further cooling increasingly tedious for the present setup. Here, an improved trap design could increase the trap lifetime, and higher detection efficiencies would make losses more tolerable. Varying the cooling cycle or choosing a molecule species with a faster spontaneous decay would allow faster cooling.

Even without these improvements, the present experiment brings key applications of cold polar molecules within reach. First, the increased phase-space density improves the sensitivity for high-resolution spectroscopy and collision experiments. Second, the obtained temperature allows loading of molecules into more weakly confining microwave~\cite{DeMille04} or optical traps which can hold molecules in their rotational ground state, a prerequisite for achieving quantum degeneracy. Finally, our chip-like trap and guide architecture matches the demands on hybrid quantum systems for quantum information processing with cold and ultracold molecules~\cite{Andre06}.

%%%%%%%%%%%%%%%%%%%%%%%%%%%%%%%%%%%%%%%%%%
\section*{Methods}

The initial sample of molecules is generated by velocity filtering~\cite{Junglen04} from a liquid-nitrogen cooled source and loaded into the electric trap~\cite{Englert11} via a quadrupole guide. Detection is performed by guiding the molecules to the QMS via a second guide. The selection of CH$_3$F as molecule species is based on favourable properties for trap loading and detection. Large rotational constants and sufficient vapour pressure down to almost $100$\,K allow efficient velocity filtering with an adequate fraction ($\sim20\,\%$) of molecules in the rotational states used for cooling. Low contributions from other molecules at the atomic mass $34$ of CH$_3$F results in very low background for the QMS detection. For dissipation, the approximately $15$\,Hz spontaneous decay rate of the most suited parallel vibrational $v_1$ symmetric C-H stretch mode is sufficient. A description of the IR and MW radiation sources required for CH$_3$F~\cite{Graner81} is provided in the appendix.

For internal-state-discriminating measurement, a microwave depletion (MWD) pulse consisting of various MW frequencies near $204$\,GHz is applied during the last four seconds before trap unloading. This mixes all $M$-sublevels of the $J=3$ and $J=4$ states, leaving only the unaffected background from molecules in other states in the trap. The difference in the unloading signal with and without the MWD applied is used for all measurements shown in Figs.~\ref{filters}--\ref{lifetime}.

The offset electric field in the centre region of the trap results in an offset Stark splitting of $180$\,MHz between neighbouring $M$-sublevels for molecules with $J=|K|=3$. For molecules with $J=4$, all Stark splittings are smaller, but this is inconsequential due to microwaves being used to couple the $|0;3,3,M\rangle$ and $|0;4,3,M+1\rangle$ states: the RF transitions will predominantly occur for $J=3$. This is particularly important for the RF filter measurement in Fig.~\ref{filters}b.

\begin{acknowledgments}
We thank P.W.H. Pinkse for support during the early stage of this experiment. Support by the Deutsche Forschungsgemeinschaft via the excellence cluster ``Munich Centre for Advanced Photonics'' is acknowledged.
\end{acknowledgments}

\bibliographystyle{unsrt}

{\color{white}end}
\appendix
\pagebreak
\begin{widetext}
\section{Detailed Cooling Scheme}\label{sec:scheme}
The illustration of the opto-electrical cooling cycle in Fig.~\ref{scheme}a in the main text includes only a simplified level scheme consisting of the molecular states and transitions which constitute the most elementary cooling cycle. Fig.~\ref{fig:scheme} shows an expanded version of the cooling cycle where we consider additional $M$-sublevels of the $J=3$ and $J=4$ states as well as additional decay channels from the $v=1$ state. We find that including the lower trapped $M$-sublevels in the cooling cycle leads to a substantial reduction in losses to untrapped $M=0$ states and that the additional spontaneous decay channels for the most part do not affect the cooling cycle.

\begin{figure*}[ht]
\centering
\includegraphics[width=.75\textwidth]{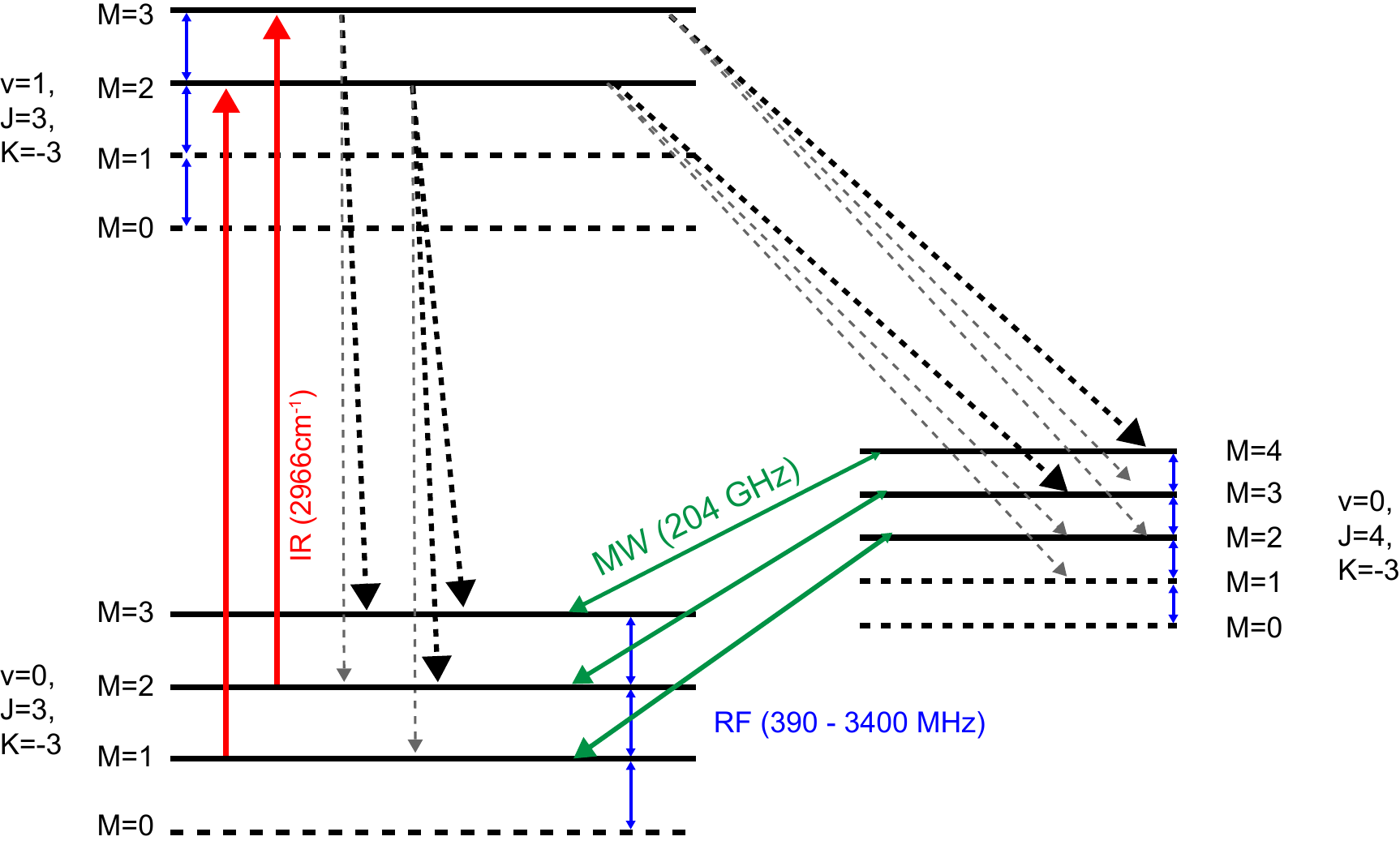}
\caption{{\bf Detailed energy-level diagram for cooling} showing the complete set of IR, MW and RF transitions as well as all spontaneous decay channels involved in cooling. Black dashed arrows indicate the key spontaneous decay channels which transfer the molecules back towards the maximally trapped states, with all other decay channels indicated by gray dashed arrows. The dashed energy levels are loss channels.}\label{fig:scheme}
\end{figure*} 

Due to the constant differential Stark splitting between neighbouring $M$-sublevels for $J$ and $K$ fixed, it is basically impossible to drive RF transitions from a state $M=J$ to $M=J-1$ (as required for cooling) without also driving transitions to lower $M$-sublevels and ultimately to untrapped $M=0$ states. For the cooling cycle, these losses are reduced by driving the RF transitions at a slow rate compared to the optical pumping back to the $M=J$ states. For an additional substantial reduction of these losses, the states $|0;3,3,1\rangle$ and $|0;4,3,2\rangle$ can be incorporated into the cooling cycle. Coupling these states via microwaves and coupling to the $|1;3,3,2\rangle$ state via the IR laser results in optical pumping back to the core states of the elementary cooling cycle, as shown in Fig.~\ref{fig:scheme}. This doubles the number of ``unlucky'' $\Delta M=-1$ RF transitions which need to occur before a molecule is lost.

Fig.~\ref{scheme}a in the main text only shows spontaneous vibrational decay to the $|0;3,3,3\rangle$ and  $|0;4,3,4\rangle$ states. Obviously, spontaneous decay to other states will also occur. However, since spontaneous vibrational decay with $\Delta J=0, \Delta M=-1$ and $\Delta J=+1, \Delta M=0$ populates states which are strongly coupled to the $v=1$ state via the IR and MW, the effect on the molecules is negligible. This leaves the decay with $\Delta J=+1, \Delta M=-1$ as the only decay channel which can induce losses. However, the Clebsch-Gordan coefficient of the $|1;3,3,3\rangle$ to $|0;4,3,2\rangle$ and the $|1;3,3,2\rangle$ to $|0;4,3,1\rangle$ transition is only $\frac{1}{144}$ and $\frac{1}{48}$, respectively. Moreover, molecules decaying to the $|0;4,3,2\rangle$ state will generally be repumped to higher $M$ states via the $|1;3,3,2\rangle$ state. Considering the few number of spontaneous decays required for cooling, losses via the $\Delta J=+1, \Delta M=-1$ spontaneous decay channel can therefore be ignored for the present experiment.

As a final consideration, we discuss two possibilities for spontaneous vibrational decay to states other than $|0;3,3,M\rangle$ and $|0;4,3,M\rangle$. First, due to the electric fields in the trap, molecular states with different $J$ are coupled so that spontaneous decay with $\Delta J>1$ in principle becomes possible. However, this was analyzed for the original proposal of opto-electrical cooling~\cite{Zeppenfeld09} where it was found that even for CF$_3$H with much smaller rotational constants, spontaneous decay with $\Delta J=2$ is marginal even for fields as large as $100$\,kV/cm. Second, it is well known that the $v_1$ C-H vibrational stretch mode of CH$_3$F is coupled to the $2v_2$ and $2v_5$ modes via a strong Fermi-resonance~\cite{Yates47,Giguere76}. As a result, spontaneous decay to the first excitation of the $v_5$ mode will occur, with a rate of about $1-2$\,Hz~\cite{Newton76}. Thus, approximately every $10$th decay will occur via this channel. The Fermi resonance thus causes a significant loss channel, evidence for which is discussed in appendix~\ref{sec:mwd}.

\section{Rate estimates for opto-electrical cooling}\label{sec:estimates}
In this section we perform some simple back-of-the-envelope type estimates for the dynamics of opto-electrical cooling. These estimates allow us to derive a scaling law for the RF losses to $M=0$ as well as to explain the discrepancy between the vibrational spontaneous decay rate of about $15$\,Hz and the cooling time of about $8$\,s per halving of the molecular temperature.

Our entire analysis of opto-electrical cooling relies on using rates to describe transitions between populations in various molecular states. We thereby ignore the possibility of any coherences, an assumption which is strongly justified by the experimental parameters. In particular, the RF transition rate of about $1$\,Hz during cooling compares to a trap frequency of about $1$\,kHz: the molecules thus come into resonance with the RF many times before a transition occurs. Similarly, the MW and IR rate of about $100$\,Hz during cooling compares to a Stark broadening of many MHz in the central region of the trap.

For the following discussion, it is useful to group the molecules into three sets of states, the strongly trapped pair of states $|0;3,3,3\rangle$ and $|0;4,3,4\rangle$, the (on average) moderately trapped set of states $|0;3,3,2\rangle$, $|0;4,3,3\rangle$, and $|1;3,3,3\rangle$, and the (on average) weakly trapped set of states $|0;3,3,1\rangle$, $|0;4,3,2\rangle$, and $|1;3,3,2\rangle$. These sets are motivated by the fact that the molecular states in each of these sets are strongly coupled to each other by IR and MW radiation. As a result, we can treat each set as a single doubly- or triply-degenerate state.

Based on the previous grouping of molecular states, cooling can be modelled as an RF which couples neighbouring sets of states with a rate of $\gamma_{\rm RF}$, and a repumping to more strongly trapped states with a rate of $\gamma_{\rm IR}$. With populations $p_{\rm s}$, $p_{\rm m}$, and $p_{\rm w}$ in the strongly, moderately, and weakly trapped set of states, respectively, and the assumption $\gamma_{\rm RF}\ll\gamma_{\rm IR}$, we find an approximate steady state for cooling with $\gamma_{\rm RF}\,p_{\rm{s}}=\gamma_{\rm IR}\,p_{\rm m}$ and $\gamma_{\rm RF}\,p_{\rm m}=\gamma_{\rm IR}\,p_{\rm w}$. We therefore find that $p_{\rm w}=p_{\rm s}\times\left(\frac{\gamma_{\rm RF}}{\gamma_{\rm IR}}\right)^2$. Since $\gamma_{\rm RF}$ is simply the rate of cooling and with losses to untrapped states given by $\gamma_{\rm RF}\,p_{\rm w}$, we find that the loss rate to untrapped states is suppressed compared to the cooling rate by a factor of $\left(\frac{\gamma_{\rm RF}}{\gamma_{\rm IR}}\right)^2$. Note that without the weakly trapped set of states, the second factor of $\frac{\gamma_{\rm RF}}{\gamma_{\rm IR}}$ would be absent and losses would scale much less favourably for small $\gamma_{\rm RF}$. Also, we have found that to avoid losses, the assumption $\gamma_{\rm RF}\ll\gamma_{\rm IR}$ is true self consistently. This is equivalent to $p_{\rm w}\ll p_{\rm m}\ll p_{\rm s}$, and we have therefore found that the majority of molecules must be in the strongly trapped states at any given time.

We now estimate realistic rates for our experiment. The spontaneous decay rate of the vibrationally excited state is approximately $15$\,Hz. However, the population in both the moderately and weakly trapped set of states is spread over three states, only one of which is in $v=1$, effectively reducing $\gamma_{\rm IR}$ by a factor three. Moreover, the branching ratio of decay from the state $|1;3,3,3\rangle$ to the strongly trapped states is about $0.75$, further reducing $\gamma_{\rm IR}$ by this factor. We therefore find $\gamma_{\rm IR}\cong4$\,Hz, and to sufficiently suppress losses, applying the RF with a rate of $\gamma_{\rm RF}=1$\,Hz seems reasonable.

To obtain the rate at which the temperature can be reduced, two additional effects play a role. First, we estimate the number of cooling cycles required to achieve a given reduction in temperature. Consider a molecule with energy $E$ in the rotational sublevel $M$. The condition for the molecule to have sufficient energy to reach resonance with an RF frequency of $f_{\rm RF}$ (allowing an RF transition to take place) is $E>M\,f_{\rm RF}$. Since the RF transition reduces the molecule's energy by a factor of $E/(E-f_{\rm RF})$, optimal cooling takes place for $f_{\rm RF}$ as large as possible. To account for a range of molecule energies, we consider $f_{\rm RF}$ such that one final RF transition is possible. For $M=3$, this is equivalent to $3\,f_{\rm RF}<E<4\,f_{\rm RF}$ which translates to an average reduction in energy by a factor of about $3.5/2.5\cong\sqrt{2}$. We thus find that approximately two cooling cycles are necessary to reduce the temperature by a factor of $2$.

The final effect for the cooling duration is that due to the stochastic nature of the cooling process, a cooling rate of $1$\,Hz means that a fraction of $e^{-1}$ of the molecules will remain uncooled after $1$\,s. The cooling rate $\gamma_{\rm RF}$ is thus substantially faster than the time required to efficiently cool molecules by one step. For a reasonable cooling efficiency of $1-e^{-3}$, this adds a further factor of $3$ to the cooling time. We therefore end up with a total of $6$\,s required to cool the molecules by a factor of $2$, which, within the accuracy of our estimate, agrees with the $8$\,s chosen for the experiment.

\section{Rate equations}\label{sec:rates}

In the previous section we showed that even a very simple estimate based on rates allows several aspects of opto-electrical cooling to be understood quantitatively. In this section, we expand our model by considering populations of molecules in the trapped states shown in Fig.~\ref{fig:scheme} and solving time-dependent rate equations describing the transitions between the various states. Due to the ability to independently determine all relevant parameters other than the absolute signal amplitude, the expanded model works particularly well for an ``accumulation'' measurement where we perform optical pumping from the moderately to the strongly trapped states. The perfect fit of the experimental data with the values expected from our model demonstrates that our model accurately describes the processes in the trap and, moreover, that accumulation works as intended. For cooling, determining all the relevant parameters for the model is not possible, resulting in only an approximate description of our data. However, the model nonetheless provides additional insight into the cooling process.

For the rate-equation model, we consider the time-dependent populations of molecules $p_i(t)$ with $i=1,...,9$ in the seven low-field seeking states $|0;3,3,M\rangle$ and $|0;4,3,M\rangle$ as well as in the two excited states $|1;3,3,3\rangle$ and $|1;3,3,2\rangle$. The state $|1;3,3,1\rangle$ can be ignored since it is strongly coupled to the untrapped $|0;3,3,0\rangle$ state via the IR laser. For a given initial state distribution $p_i(0)$, we determine the state evolution by solving the set of rate equations $\dot{p_i}(t)=\sum_jA_{i,j}p_j(t)$. Here, $A_{i,j}$ is the rate at which molecules in the state $j$ are transferred to the state $i$, describing the effect of the radiation fields which are applied during the experiment as well as the spontaneous decay from the $v=1$ states.

\subsection{Accumulation}
Our rate-equation model can be applied particularly successfully to an accumulation measurement where the molecules are optically pumped from the moderately trapped states $|0;3,3,2\rangle$ and $|0;4,3,3\rangle$ to the strongly trapped states $|0;3,3,3\rangle$ and $|0;4,3,4\rangle$. Experimentally, we start by performing the cooling sequence to $1.22$\,GHz as described in the main text. This is followed by two seconds during which microwaves are applied to couple the four states $|0;4,3,4\rangle$, $|0;3,3,3\rangle$, $|0;4,3,3\rangle$, and $|0;3,3,2\rangle$, thereby distributing the molecules among these four states. During the subsequent four seconds, we apply the MW and IR fields shown in Fig.~\ref{fig:scheme} (but no RF), identical to the accumulation as described in the main paper, for a variable amount of time $T$. This causes a certain fraction of the molecules to be repumped to the strongly trapped states. After the four seconds, we reduce the confining electric field strength of the trap and unload the molecules from the trap for detection, with the molecular state affecting the probability for detection.

\begin{figure}[t]
\centering
\includegraphics{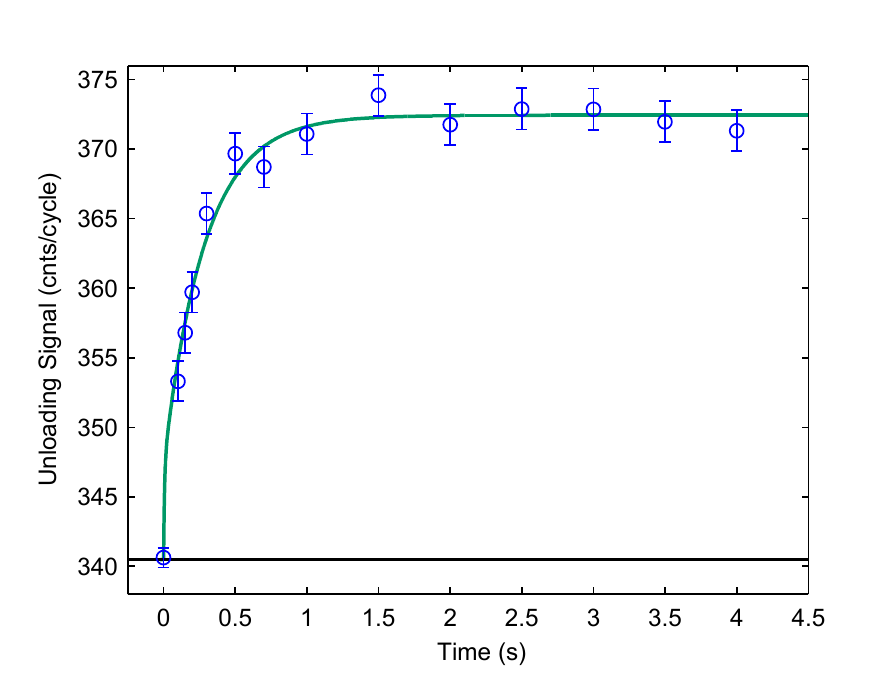}
\caption{{\bf Accumulation.} We show the total integrated QMS signal during unloading as a function of the time during which the radiation fields for accumulation are applied. The solid line is the expected time dependence of the signal from the rate-equation model, showing excellent agreement with the experimental data.}\label{fig:accum}
\end{figure} 

The accumulation measurement is particularly suited for comparison to the rate-equation model since all aspects of the experiment not included in the model can be accounted for. In particular, the energy distribution of the molecules in the trap is not included in the model. The energy of a molecule strongly affects its detection efficiency and can in principle affect the transition rates between the molecular states. However, due to the box-like shape of our trap potential~\cite{Englert11}, the probability for a molecule to be in the central trap region is close to unity and, in particular, rather insensitive to a molecule's energy. Hence, the rate of the MW and IR transitions, which occur in the central trap region, will also be quite insensitive to a molecule's energy. As a result, we expect the two seconds of mixing to create similar energy distributions for the molecules in each of the molecular states and for this to remain the case throughout the accumulation. Moreover, we expect the energy distribution in each of the final states to be independent of the duration of the accumulation. Consequently, we can define energy-averaged detection probabilities $d_{\rm s}$ and $d_{\rm m}$ for molecules in the strongly and moderately trapped states, respectively, which are independent of the accumulation time $T$.

Based on the previous discussion, the detected signal for the accumulation measurement is composed of individual contributions as follows. Molecules which are in the strongly trapped states at the start of the accumulation are unaffected by the accumulation and, together with the background count rate, will provide a constant contribution $S_{\rm s}$ to the detected signal. For molecules initially in the moderately trapped states, a fraction $p(T)$ depending on the duration of the accumulation ends up in the strongly trapped states and the rest remains in the moderately trapped states. For $N_{\rm m}$ molecules initially in the moderately trapped states, this results in a contribution $N_{\rm m}\,d_{\rm s}\,p(T)+N_{\rm m}\,d_{\rm m}(1-p(T))$ to the detected signal. As a result, the total signal is of the form $a+b\,p(T)$, where $a$ and $b$ are taken as fit parameters. Since the MW and IR transition rates as well as the spontaneous decay rate are known from other measurements, the function $p(T)$ can be precisely calculated, so that the vertical scaling and vertical position remarkably remain the only fit parameters.

The experimental data and the expected time dependence obtained from the rate-equation model is shown in Fig.~\ref{fig:accum}. The value of chi-squared for the variation of the data around our model is $\chi^2=7.4$, which, for our $\nu=12$ degrees of freedom results in $\chi^2/\nu=0.617$, showing that the model and our data are in perfect agreement. Note that after about $1$\,s, the signal no longer increases. This is due to all molecules having been pumped to the strongly trapped states by this time. While accumulation from the weakly trapped to the strongly trapped state is a bit slower, $1$\,s is still sufficient to transfer about $84$\,\% to the strongly trapped states. This proves that after $1$\,s of accumulation, practically all molecules in the states $|0;3,3,M\rangle$ and $|0;4,3,M\rangle$ are in the top $M$-sublevels.

\begin{figure}[t]
\centering
\includegraphics{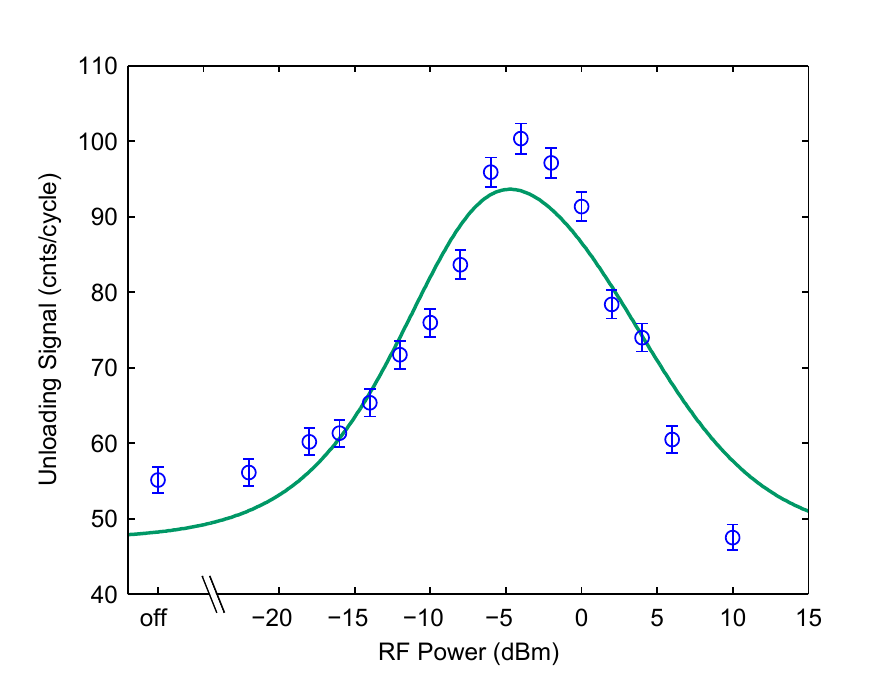}
\caption{{\bf Cooling efficiency vs.\ RF power.} We show the number of detected molecules as a function of the power of the $390$\,MHz RF during the last $4$\,s of cooling. The solid line is the expected dependence of the signal from the rate-equation model.}\label{fig:rf}
\end{figure} 

Finally, we explain the extremely steep initial increase of the signal. This is due to the strong MW and IR fields coupling the three moderately trapped states. As a result, even for very short accumulation duration, one third of the molecules initially in the $v=0$ moderately trapped states are transferred to the state $|1;3,3,3\rangle$. These molecules then have four seconds to decay back to the $v=0$ states, with a large fraction ending up in the strongly trapped states.

\subsection{Cooling}
We now apply the rate-equation model to opto-electrical cooling. This is considerably more difficult since the energy of the molecules now plays a substantial role. In particular, the RF transitions not only change the energy of the molecules but also occur at a rate strongly dependent on the energy, forcing us to keep track of the molecules' energy in addition to their internal state. One approach to take into account the energy of the molecules is to switch from discrete state populations $p_i$ to energy distributions for each of the molecular states $p_i(E)$. This approach can be simplified considerably by making the following approximation. Referring to appendix~\ref{sec:estimates}, we approximate the RF rate to be a constant times the relevant Clebsch-Gordan coefficient for $E>M\,f_{\rm RF}$ and zero otherwise. This allows the energy of the molecules to be discretized: $p_{i,n}$ is the population of molecules in state $i$ with energy $n\,f_{\rm RF}<E<(n+1)f_{\rm RF}$. RF transitions only couple populations $p_{i,n}$ with $n\ge M_i$ where $M_i$ is the $M$ quantum number of the state $i$. Moreover, each RF transition changes $n$ by $1$.

Finding a suitable cooling measurement for comparison to our model is considerably more difficult than for accumulation due to the inability to measure state-dependent energy distributions of the molecules sufficiently accurately. We therefore perform an indirect measurement, the number of molecules cooled as a function of the RF power. Experimentally, we perform the full cooling sequence to $390$\,MHz. However, the RF power of the final RF at $390$\,MHz is varied. This is followed by a $1$\,s RF knife-edge filter at $390$\,MHz to eliminate molecules which have not been cooled. The detected signal as a function of the final RF power for cooling is shown in Fig.~\ref{fig:rf}.

As discussed in appendix~\ref{sec:estimates}, each $4$\,s interval in the cooling sequence is expected to result in approximately one cooling cycle. As a result, we chose the initial state distribution for the model to consist of molecules in the strongly trapped states with $2\,f_{\rm RF}<E<4\,f_{\rm RF}$. The population with $E<3\,f_{\rm RF}$ is unaffected by the RF and provides a constant background. The rest of the population can be cooled by one additional cycle. For comparison with the experimental data, we show the calculated number of molecules with $E<M\,f_{\rm RF}$ after $4$\,s of cooling as a function of the RF power in Fig.~\ref{fig:rf}. Such molecules are no longer affected by the RF and would therefore also be unaffected by the knife-edge filter. As fit parameters we have the initial number of molecules with $E<3\,f_{\rm RF}$ and $E>3\,f_{\rm RF}$ as well as a scaling factor for the RF power.

\begin{figure}[t]
\centering
\includegraphics{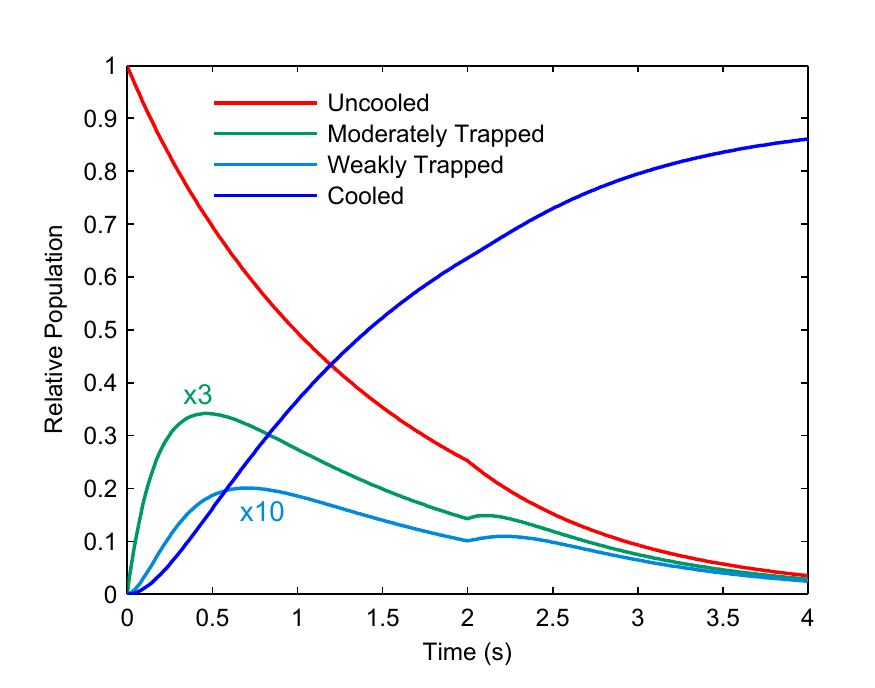}
\caption{{\bf State distribution during cooling.} We show the fraction of molecules in the strongly trapped states remaining at the initial energy, the fraction of molecules in the moderately and weakly trapped states whose energy has been reduced by $f_{\rm RF}$ and $2f_{\rm RF}$, respectively, and the fraction of molecules which have experienced at least one spontaneous decay to a more strongly trapped set of states, and have therefore been cooled. Note that the populations in the moderately and weakly trapped states have been scaled by a factor of $3$ and $10$, respectively.}\label{fig:states}
\end{figure} 

Although the agreement between the model and the experimental data is by no means perfect, the model and the experimental data show the same key features. In particular, the signal strongly decreases for both too much as well as too little RF power. The decrease in signal for low power is a key signature of our cooling process: as the RF power is ramped up, RF transitions to lower $M$-sublevels start to occur, but the IR rapidly pumps these molecules back to the top $M$-sublevels leading to cooling. As a result, the molecules no longer have enough energy to be pumped to untrapped $M=0$ states by the much stronger RF knife after the cooling sequence, resulting in a signal increase. The large factor $2$ signal difference between optimal RF power and no/too much RF power is remarkable, indicating the large number of molecules which perform at least one cooling cycle. Note that similar measurements for all other RF frequencies in the cooling sequence result in a similar signal increase, and have in fact been used to determine the optimal RF power for all RF frequencies.

As a final application of the rate-equation model to cooling, we show the calculated time evolution of the population of molecules in the various molecular states in Fig.~\ref{fig:states}. We again start with all molecules in the strongly trapped states, but with energy $3\,f_{\rm RF}<E<4\,f_{\rm RF}$, allowing one final cooling cycle. Note that for all cooling measurements where constant RF cooling frequencies are applied for $4$\,s at a time, we increase the RF power by $2$\,dB after $2$\,s to increase the cooling efficiency. Similarly, we here increase the RF rate by $2$\,dB after $2$\,s resulting in the kinks at $t=2$\,s. The initial RF rate is chosen such that the largest number of molecules has been cooled after $4$\,s.

The main result of Fig.~\ref{fig:states} is to confirm that throughout cooling, almost all the molecules remain in the strongly trapped states, with the relative population in other states never exceeding $15$\,\% and with the population substantially lower by the end of the $4$\,s of cooling. Even taking into account the simple nature of our model, we can therefore expect the great majority of molecules in the states $|0;3,3,M\rangle$ and $|0;4,3,M\rangle$ to end up in the top $M$-sublevels for all measurements shown in the main paper except no-cooling.

Despite the small fraction of molecules in the less strongly trapped states, the calculations for Fig.~\ref{fig:states} still show that approximately $9$\% of molecules are lost to untrapped states. We compare this value to the losses observed during cooling. A $9$\,\% reduction in molecules every $4$\,s results in a factor $1.46$ fewer molecules during the final $16$\,s of the cooling cycle. Together with the losses due to the finite trap lifetime, these losses account for a factor $2.8$ fewer molecules for the cooling cycle to $390$\,MHz compared to the cooling cycle to $1220$\,MHz, which is practically the same as the losses observed in the experiment. The losses predicted by the rate-equation model are thus consistent with the losses observed during opto-electrical cooling.

\section{Internal-State-Discriminating Measurement}\label{sec:mwd}
In the experiment, molecules are detected with a quadrupole mass spectrometer which is insensitive to the internal molecular state. We therefore perform microwave depletion (MWD) measurements to eliminate molecules in the rotational states involved in the cooling, allowing the remaining background from molecules in other rotational states to be subtracted. After a detailed description of the MWD, we show the results from a MWD saturation measurement to prove the effectiveness of the MWD. The nonzero saturation level suggests the existence of decay channels from the vibrationally excited state to states other than $J=3$ and $J=4$. We also discuss the contribution of molecules with $|K|=1$ and $|K|=2$ to the MWD signal.

\begin{figure}[t]
\centering
\includegraphics[width=1\textwidth]{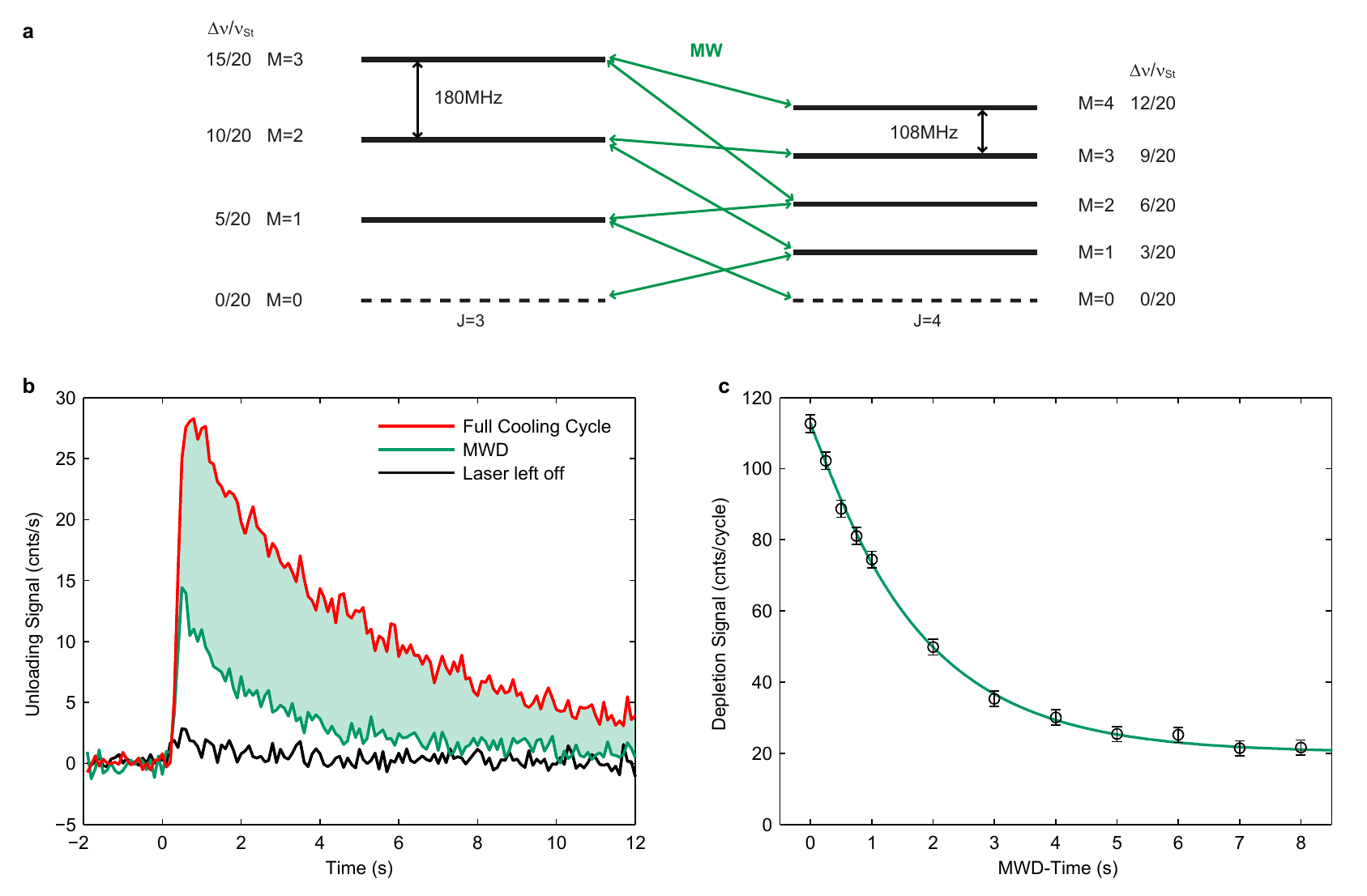}
\caption{{\bf Microwave Depletion.} {\bf a,} Set of microwave transitions applied to the molecules during the MWD pulse to remove molecules in the rotational states $|0;3,3,M\rangle$ to $|0;4,3,M\rangle$ from the trap. $\Delta \nu /\nu_{\rm{St}} = \frac{KM}{J(J+1)}$ denotes the Stark detuning of the rotational states in the homogeneous-field central region of the electric trap relative to the Stark shift $\nu_{\rm{St}}=\left|{\mathbf E}\right|\cdot\left|{\mathbf d_{\rm{el}}}\right|\cong720$\,MHz of a maximally aligned molecule. Note that transitions with $\Delta M=\pm 1$ are favoured since the microwaves are expected to be polarized perpendicular to the electric field in the center of the trap. {\bf b,} Unloading signal after the complete cooling cycle as well as for the same cycle but with either the IR laser left off or the MWD applied for $4$\,s. The shaded region represents the state-discriminated unloading signal. {\bf c,} Variation of the remaining molecule signal as a function of the MWD pulse duration. The MWD pulse is applied during a fixed $8$\,s of storage after the complete cooling cycle. The solid curve shows the predicted dependence based on rate equations.}\label{fig:mwd}
\end{figure}

For MWD, we apply a MW pulse to the molecules to drive all $\Delta M=\pm1$ transitions between the states $|0;3,3,M\rangle$ and $|0;4,3,M\rangle$ with $M\ge0$, as shown in Fig.~\ref{fig:mwd}a. This mixes the populations in all these states. Molecules which reach the $M=0$ states are lost from the trap practically instantaneously. The effect of the MWD on the trap unloading signal is shown in Fig.~\ref{fig:mwd}b. The difference in signal with and without the MWD constitutes the state-discriminated unloading signal, used for the measurements in the main paper.

Fig.~\ref{fig:mwd}c shows the number of molecules surviving the MWD pulse as a function of the duration of the pulse. A MWD pulse duration of $4\,$s, used for the MWD measurements in the main paper, is clearly sufficient to obtain strong saturation of the depletion. Since the relative transition rates for all the MW transitions are known from the Clebsch-Gordan coefficients, the differential Stark shifts, and the relative MW intensities, the rate-equation model from appendix~\ref{sec:rates} should accurately describe the data, predicting the solid curve in Fig.~\ref{fig:mwd}c. The model allows us to extract absolute transition rates for all microwaves applied in the experiment.

In principle, MWD can also be used to resolve the $M$-sublevels of the rotational states by driving only a subset of the transitions shown in Fig.~\ref{fig:mwd}a. However, attempts to be $M$-resolved suffer from Stark broadening which prevents perfect separation of the MW transitions. In particular, the $|0;3,3,M\rangle$ to $|0;4,3,M-1\rangle$ transitions have to be driven with high MW power due to both their small Clebsch-Gordan coefficients and their high differential Stark shifts. Each of these transitions therefore inevitably also drives other MW transitions in electric fields outside the trap centre, precluding simple $M$-resolved measurement. However, the results from appendix~\ref{sec:rates} prove that after both cooling and accumulation, the molecules are predominantly in the strongly trapped states $|0;3,3,3\rangle$ and $|0;4,3,4\rangle$.

Despite clear saturation of the MWD, a substantial fraction of the molecules remain unaffected by the MWD even after $8$\,s. This is surprising considering the laser-off signal in Fig.~\ref{fig:mwd}b is substantially lower than the signal with the MWD and the laser applied: the remaining molecules must have been addressed by the laser at some point. Considering that the $|0;3,3,M\rangle$ to $|1;3,3,M\rangle$ transition is relatively isolated in the $v_1$ vibrational band, the most plausible explanation for the remaining molecules is that they have decayed to an excited vibrational $v_5$ state. This is possible due to the strong Fermi-resonance between $v_1$ and $2v_5$, as discussed at the end of appendix~\ref{sec:scheme}. However, precisely determining which states the remaining molecules occupy would require new MW or IR sources.

Although the MWD pulse is optimized for $|K|=3$, the zero-field frequencies of the $|0;3,1,M\rangle$ to $|0;4,1,M\rangle$ and $|0;3,2,M\rangle$ to $|0;4,2,M\rangle$ transitions are less than $30\,$MHz detuned from the zero-field frequency of the $|0;3,3,M\rangle$ to $|0;4,3,M\rangle$ transition. As a result, the MWD pulse also drives these transitions, and molecules in the states $J=3$ and $J=4$ with $|K|=1$ and $|K|=2$ will also contribute to the state-discriminated signal. This effect will be largest for the no-cooling reference measurement since the population of molecules in the states with $|K|=3$ has not yet been enhanced by accumulation or cooling. Nonetheless, the fraction of molecules loaded into the trap in the states with $J=3,4$ which are in the states $|K|=1,2$ is only about $25$\,\% due to reduced Stark shifts and half the hyperfine degeneracy. The density measured for the no-cooling is therefore overestimated by only a small amount.

For the accumulation and cooling measurements, the contribution to the state-discriminated signal of molecules with $|K|=1,2$ is considerably reduced. In particular, the zero-field $\Delta J=0$ IR transitions from the states $|0;3,1,M\rangle$, $|0;4,1,M\rangle$, $|0;3,2,M\rangle$, and $|0;4,2,M\rangle$ are all $4\,$GHz or more blue detuned from the $|0;3,3,M\rangle$ to $|1;3,3,M\rangle$ transition~\cite{Graner81}, and the laser is therefore red detuned by a similar amount. As a result, the only IR transitions from these states which can occur, if they occur at all, are $\Delta M=-1$ transitions, resulting in optical pumping to untrapped $M=0$ states. From this we see that, compared to no-cooling, the absolute number of $|K|=1,2$ molecules in the accumulation signal will, if anything, decrease and the relative number is reduced by at least a factor of $2$ due to the increased number of molecules with $|K|=3$. For the cooling measurements, the RF fields will lead to strong losses for molecules with $|K|=1,2$, strongly reducing their absolute number. In fact, for the complete cooling cycle to $390\,$MHz, the molecules with $|K|=1,2$ must be entirely contained in the ``laser-off'' signal, and since this is practically zero, the state-discriminated signal essentially only includes molecules with $|K|=3$. 

%\FloatBarrier
\section{Details of the Experiment} 

IR radiation at $2966$\,cm$^{-1}$ to drive the $J=3$, $\Delta J=0$ transition of the $v_1$ vibrational band is produced by a CW optical parametric oscillator locked to a frequency comb. Line assignments and frequency values published in Ref.~\cite{Graner81} are verified with saturation spectroscopy with sub-MHz accuracy using a multi-pass Herriot cell and validated via combination differences between the $|v;J,K\rangle$ and $|v;J+1,K\rangle$ states. The $410$\,mW of IR power illuminates the trap from the side (as shown in Fig.~\ref{scheme}b in the main text), driving the $\Delta M=+1, \Delta J=0$ transitions shown in Fig.~\ref{fig:scheme} with an estimated rate on the order of $100$\,Hz.

To couple the $v=0$, $J=3$ and $J=4$ states, MW radiation at $204.24$\,GHz is generated by duodectupling (x12) the output of a frequency synthesizer. On the order of $5$\,mW MW power radiate from a MW horn antenna onto the front of the trap. During the cooling cycle the effective power is reduced by applying the MW with an overall duty cycle of $25\%$, providing sufficient MW intensity to alternatingly drive the three narrowband $J=3$ to $J=4$ transitions with an approximate rate of $100\,$Hz. For efficient MW depletion we require the full $100\%$ duty cycle.

RF to couple neighbouring rotational $M$-sublevels and to apply the RF knife filter is applied directly to the microstructure electrode contact leads. Naturally occurring electric resonances are exploited to inject sufficient RF power.

Due to the varying electric fields in an electric trap, transitions between various $M$-sublevels will be addressed by the same MW or IR frequency at different locations in the trap, potentially driving unwanted transitions to untrapped states. The choice of transitions shown in Fig.~\ref{fig:scheme} eliminates this problem for the IR and strongly reduces it for the MW. In particular, for the IR transitions, the $M$-sublevels of a given rotational state have very similar Stark shifts for both the $v=0$ and the $v=1$ state. In non-zero fields, all $\Delta M=-1$ transitions occur at a red-detuned frequency compared to the zero-field frequency, the $\Delta M=0$ transitions occur near the zero-field frequency, and the $\Delta M=+1$ transitions are blue detuned. We can therefore easily avoid the $\Delta M=-1$ and $\Delta M=0$ transitions while driving all transitions between the states $|0;3,3,M\rangle$ and $|1;3,3,M+1\rangle$ simultaneously by blue detuning the IR laser from the zero-field frequency. 

Driving the three MW transitions from $|0;3,3,M\rangle$ to $|0;4,3,M+1\rangle$ shown in Fig.~\ref{fig:scheme} while avoiding unwanted MW transitions turns out to be particularly easy. Due to the small differential Stark shifts between these pairs of states as well as the relatively large Clebsch-Gordan coefficients, relatively low MW power is sufficient to drive these transitions. Due to the substantially larger differential Stark shifts of the unwanted transitions, the low MW power and the homogeneous electric fields in the trap cause the unwanted transitions to occur at a sufficiently slow rate.

Due to low detection efficiency, extracting a velocity distribution using a time-of-flight measurement requires an optimized data acquisition strategy to increase the signal per cycle. Therefore, instead of using only the first rising edge of the unloading signal, the quadrupole guide to the QMS is repeatedly switched on and off during each trap unloading. This allows several time-of-flight profiles (rising edges) to be recorded each experimental cycle which are summed up. In addition to improving the data acquisition rate, this results in a rising edge which represents an average over the velocity distribution of molecules leaving the trap at different times after the unloading process has started.

To determine the density of the molecular ensemble in the trap, a density calibration measurement has been performed. This calibration is necessary since the molecular cloud is not measured directly in the trap but transferred to the ionization volume of the QMS via a quadrupole guide, which samples the molecule density in the trap. The flux of molecules from the quadrupole guide is obtained by integrating the density of molecules measured by the QMS over the transverse position of the QMS and multiplying with the velocity of the molecules. A lower bound for the number of molecules in the trap is then obtained by integrating the flux over time during unloading. This value assumes that all molecule losses occur via the exit guide. Accurate relative values for the density in the trap can be obtained by comparing the initial molecule density at the QMS with the size of the exit hole from the trap, which is proportional to $\frac{T^2}{E^2\langle\cos\theta\rangle^2}$. $T$ is the temperature of the molecules, $E$ is the unloading electric field strength, and $\langle\cos\theta\rangle$ is the state-dependent orientation of the molecules.

\section{Unloading Electric Field Strength}

\begin{figure}[hbt]
\centering
\includegraphics{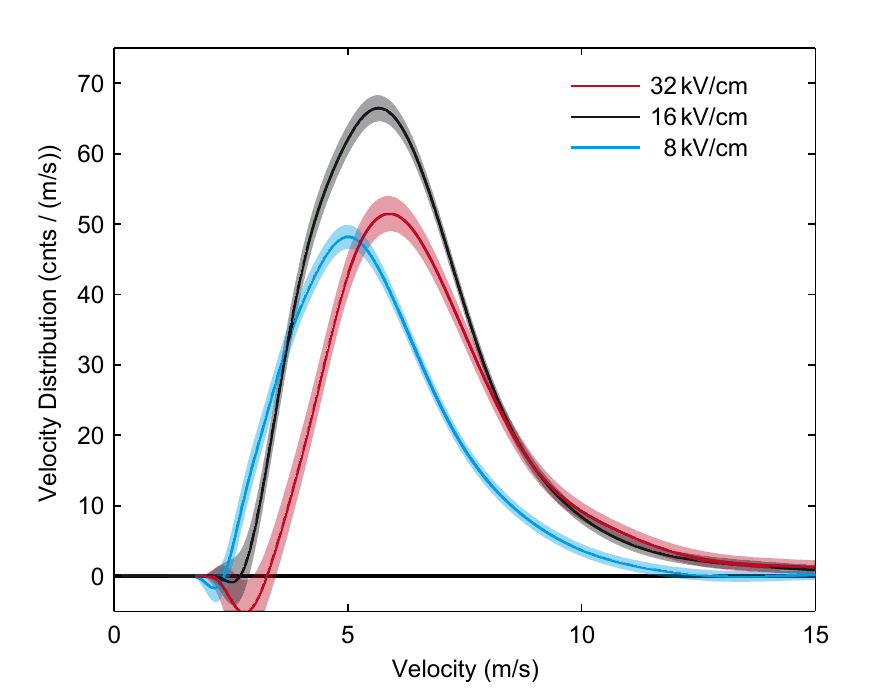}
\caption{{\bf Effect of the unloading electric fields on the measured velocity distribution.} We show the velocity distribution for cooling to $1.22$\,GHz ($16\,$kV/cm) from the main paper as well as the distribution for both half ($8\,$kV/cm) and twice ($32\,$kV/cm) the unloading electric field strength. The correct unloading electric field results in a measured velocity distribution representing practically all molecules in the trap.}\label{fig:unloading}
\end{figure}

As already discussed in the main text, the unloading electric field strength plays a key role in properly measuring the velocity distribution of the molecules in the trap. Specifically, slower molecules are not efficiently extracted from the electric trap at high unloading fields whereas faster molecules are lost from the experiment at lower unloading fields due to weaker confinement. To investigate this effect, we have measured the velocity distribution for cooling to $1.22$\,GHz for a variety of unloading fields, with three representative distributions shown in Fig.~\ref{fig:unloading}. As expected, an unloading field which is too low results in a strong decrease in fast molecules, and an unloading field which is too high reduces the number of slow molecules. The unloading field of $16\,$kV/cm with the maximal signal, chosen for the main paper, results in a velocity distribution which encompasses the other two distributions almost entirely, demonstrating that this field strength has been suitably chosen. For cooling to $390$\,MHz, the result of varying the unloading field about the optimum is practically the same as for $1.22$\,GHz. For accumulation, the increase in slower molecules for lower fields and the increase in faster molecules for higher fields is slightly larger, with the behaviour otherwise similar, indicating a slightly wider velocity distribution in the trap which is difficult to measure with a single unloading voltage. We emphasize that in all three cases, the unloading field has been varied about the optimal values, which vary by over a factor of $10$, and not about a fixed value. The present measurement thus confirms the necessity to adjust the unloading fields according to the temperature of the molecules.
\color{white}
\end{widetext}

\end{document}